\documentclass[a4paper,fleqn,usenatbib]{mnras}

\usepackage{newtxtext,newtxmath}

\usepackage[T1]{fontenc}
\usepackage{ae,aecompl}

\usepackage{graphicx}	
\usepackage{amsmath}	
\usepackage{amssymb}	

\title[Almost periodic functions and Blazhko stars]
{On the connection between almost periodic functions and Blazhko light curves}

\author[J. M. Benk\H{o}]{
J\'ozsef M. Benk\H{o},$^{1}$\thanks{E-mail: benko@konkoly.hu}
\\
$^{1}$Konkoly Observatory, MTA CSFK, Konkoly Thege M. u. 15-17., H-1121 Budapest, Hungary\\
}

\date{Accepted 2017 September 5.  Received 2017 September 4; in original form 2017 July 10}

\pubyear{2017}

\begin{document}
\label{firstpage}
\pagerange{\pageref{firstpage}--\pageref{lastpage}}
\maketitle

\begin{abstract}
In this paper it is shown that the mathematical form which  
most precisely describes the Blazhko RR\,Lyrae light curves 
is connected to almost periodic functions and not
to the mathematics of modulation. That is, the Blazhko effect is more
than a simple external modulation of the pulsation signal. 
The mathematical framework of almost periodic functions predicts a new observable effect:
a shift of the Fourier harmonics of the main pulsation frequency from the exact harmonic
position. This phenomenon is called as harmonic detuning effect (HDE).
The published deviations of the harmonics of V445\,Lyr are explained with this effect.
 HDE is also found for V2178\,Cyg, another Blazhko star observed by 
the {\it Kepler} space telescope. 
HDE is detectable only if the phase variation part of the Blazhko effect is of large amplitude and 
non-periodic enough, additionally, the time span of the observed light curve is sufficiently 
long for obtaining precise frequencies. These three conditions restrict the number of stars showing 
detectable HDE and explain why this effect has not been discovered up to now. 

\end{abstract}




\section{Introduction}

The recent photometric space missions {\it Kepler} \citep{Borucki10} and 
CoRoT \citep{Baglin06} served us long and precise time series 
for huge amount of variable stars, including RR Lyrae stars showing Blazhko effect.
Even the definition of the effect has been refined using space photometry 
\citep{Benko10, Szabo14, Benko14, Benko16}, namely, that the Blazhko effect means a simultaneous
amplitude and frequency variation of the RR Lyrae light curves with the same period(s).

From  a mathematical point of view the Blazhko RR Lyrae light curves are described either as modulated
signals (\citealt{Benko2011} and references therein) or signals of a beating phenomenon 
(\citealt{Kolenberg06} and its references). As is shown in  
\citet{Benko2011} the modulation framework can explain many but not all 
observed light curve and Fourier spectrum features.
The formulae of the modulation picture were applied to the {\it Kepler} observations of
V445 Lyr, a Blazhko star with a strong effect, and give only 
a moderately good fit \citep{Guggenberger2012}.
The large fitting residual was explained by the influence of the  
low-amplitude frequencies and the irregular nature of the star's Blazhko effect. 
However, \citet{Szeidl2012} demonstrated that the modulation formulae result 
in an inappropriate fit for CM\,UMa as well, although the star shows 
only a simple sinusoidal amplitude variation. Then they suggested a
complex formula which fits the light curve properly.

This paper investigates the problem a bit further. As we will see, the solution is to reject
the simple modulation framework and to apply a more complex description using almost 
periodic functions. Beyond the mathematical significance of this new framework it predicts
a new observable effect and some physical consequences, as well. 
The potential usability of almost periodic functions has already been raised 
in two previous conference presentations \citep{Hakone_conf, Visegrad_conf}.
Present article discusses their actual use in detail.

\section{Motivation}

The light curves of the RR\,Lyrae stars 
are comprehended as non-sinusoidal periodic signals represented by their Fourier sums.
Following \citet{Benko2011} a generally non-sinusoidally modulated RR Lyrae light curve
can be described as:
\begin{multline}\label{mod_nsinC} 
m(t)=\left[ a^{\mathrm {A}}_0+ g^{\mathrm {A}}(t) \right] 
\Bigg\{ a_{0} +  \\
\left. \sum_{i=1}^n a_i \sin \left[ 2\pi i f_0 t 
+ \varphi_i + i g^{\mathrm {F}}(t) \right] \right\},
\end{multline} 
where the modulation functions are
\begin{equation}
 g^{\mathrm M}(t)=\sum_{j=1}^{l^{\mathrm M}}{a_{j}^{\mathrm M}\sin (2\pi j f_{\mathrm m} t 
+ \varphi_{j}^{\mathrm M}}), \ \ \ {\mathrm {M = A, \ or \ F}}.
\end{equation}
Here $f_0$ and $f_{\mathrm m}$ mean the main pulsation and the
modulation frequencies, $i, j$ are integer running indices, 
$a$ and $\varphi$ coefficients are the Fourier amplitudes and phases, respectively.
The $a^{\mathrm {A}}_0$, $a_{0}$ are the zero point constants; $n$ and $l^{\mathrm M}$
integers show the number of terms in the finite Fourier sums. 
The upper index A indicates the amplitude modulation (AM) and index F means
the frequency modulation (FM). 
From now on the independent variable $t$ is referred as `time', because this paper
deals with time dependent functions only.
The formula (\ref{mod_nsinC}) can be rewritten as:
\begin{multline}\label{mod_nsinC_rw}
m(t)=a^{\mathrm {A}}_0 a_{0} + a_{0} g^{\mathrm {A}}(t) + \\
\sum_{i=1}^n { \left[ a^{\mathrm {A}}_0 a_i + a_i g^{\mathrm {A}}(t) \right] \sin \left[ 2\pi i f_0 t
+ \varphi_i + i g^{\mathrm {F}}(t) \right]}.
\end{multline}

If we do not assume anything about the light curves,
we can start with the observed Fourier spectra as well. 
This was exactly what \citet{Szeidl2012} did. They searched for a closed form
which fits the best the observed light curves. 
They received complicated $g^{\mathrm{M}}(t)$ modulation 
functions which depend on the  Fourier orders $i$ of the carrier wave:
\begin{equation}
 g^{\mathrm M}_i(t)=\sum_{j=1}^{l_i^{\mathrm M}}{a_{ij}^{\mathrm M}\sin (2\pi j f_{\mathrm m} t 
+ \varphi_{ij}^{\mathrm M}}), \ \ \ {\mathrm {M = A \ or \ F}}.
\end{equation}
Their entire light curve fitting formula is
\begin{multline}\label{mod_nsinC2}
m^{*}(t)=m_0+\sum_{k=1}^l {b_k\sin(2\pi kf_{\mathrm m} t +\varphi^{\mathrm b}_k)}+ \\
\sum_{i=1}^n{ \left[a_i+g^{\mathrm A}_i(t)\right]
\sin\left[ 2\pi i f_0 t + \varphi_i + g^{\mathrm F}_i(t) \right]}.  
\end{multline}
Comparing this formula with Eq.~(\ref{mod_nsinC_rw}) we see an additional difference
beyond the different modulation
functions: the second term in Eq.~(\ref{mod_nsinC2}) describes an independent zero point 
(a.k.a average brightness) variation, while in the formula (\ref{mod_nsinC_rw}) this
variation directly depends on the amplitude modulation function $g^{\mathrm{A}}(t)$.  

I emphasize that formula (\ref{mod_nsinC2}) was found on empirical
basis: this format fits better the 
observed light curves than Eq.~(\ref{mod_nsinC_rw}), however, 
there was no explanation why the modulation functions depend on the 
harmonic orders of the carrier wave's Fourier solution. These  
 Fourier harmonics have only mathematical meaning. They represent
the non-sinusoidal nature of the light curves but they have no physical meaning. How 
is it possible to have a modulation in the star which  influences different ways for 
each harmonic?  

\section{Almost periodic functions}

To answer this question, first, let us reproduce some basic definitions.
An $x(t)$ real function is periodic with the period $P$ if
\begin{equation}\label{def_per}
x(t)=x(t+P), \ \ {\mathrm {or}} \ \  \vert x(t) - x(t+P) \vert = 0.
\end{equation}
As it is well-known $x(t)$ function has a unique Fourier representation which
can be written in different forms. One of them is as follows:
\begin{equation}\label{F0}
x(t)=\frac{a_0}{2} +
\sum_{n=1}^{\infty}{a_n \sin \left( 2\pi n f_0 t + \varphi_n \right)},
\end{equation}
where $a_0$ is the zero point, $f_0=1/P$ is the frequency, 
$a_n$, and $\varphi_n$ are the usual Fourier coefficients, 
$n$ is an integer index.

Now I define an extension of the real periodic functions.  
Harald Bohr was the first who defined almost periodic functions  
in the 1920s. 
%
By following his definitions \citep{Bohr1947}:
`Let us take an arbitrary function $z(t)$ continuous for $-\infty < t < \infty$.
The real number $\tilde{P}$ will then be called a translation number of $z(t)$
corresponding to $\varepsilon$ whenever
\begin{equation}\label{Qp}
\vert z(t) - z(t+\tilde{P}) \vert \le \varepsilon
\end{equation} 
(\dots) A $z(t)$ function will be
called almost periodic when, (\dots) to every $\varepsilon$>0, a length ${L}={L}(\varepsilon)$
of some sort exists such that each interval of length  $L(\varepsilon)$
contains at least one translation number $\tilde{P}=\tilde{P}(\varepsilon)$.'
Comparing the formula (\ref{Qp}) with (\ref{def_per}) we see that if $\varepsilon \rightarrow 0$
almost periodic function $z(t)$ became fully periodic and  
the translation number plays the role of period.

Almost periodic functions have many similar properties to 
the periodic ones: e.g. they form a complete orthogonal system,
or they have unique Fourier series \citep{Bohr1947}. 
The Fourier representation of a $z(t)$ function in 
the most suitable form for us now is
\begin{equation}\label{F1}
z(t)=\frac{a_0(t)}{2} +
\sum_{n=1}^{\infty}{a_n(t) \sin \left[ 2\pi n \int_0^{t} f_0 (\tau)d
\tau + \varphi_n(t) \right]}.
\end{equation}
Here,  as opposed to the expression (\ref{F0}),
all $a_0(t)$, $f_0(t)$, $a_n(t)$, and $\varphi_n(t)$ quantities are 
(time-dependent) functions.

If we compare Eq.~(\ref{mod_nsinC_rw}) and (\ref{mod_nsinC2}) with Eq.~(\ref{F1}) we see that  
both expressions (\ref{mod_nsinC_rw}) and (\ref{mod_nsinC2}) show Fourier representation of
almost periodic functions. It is easily seen that the zero point variation 
function ($a_0(t)/2$) could either be the first two terms of (\ref{mod_nsinC_rw})
or  (\ref{mod_nsinC2}). Similarly, the amplitude variation function $a_n(t)$ can
correspond to the coefficients of the  $\sin$ functions in the square brackets 
of the third term. The $\varphi_n(t)$  function can be equated both 
for Eq.~(\ref{mod_nsinC_rw}) and (\ref{mod_nsinC2}) by the last two terms in 
the argument of their $\sin$ functions. By examining the frequency variation functions
we can see that both the modulation picture (\ref{mod_nsinC_rw}) and the
best-fitting mathematical model  (\ref{mod_nsinC2}) assume fixed fundamental
frequency since
\begin{equation}
\int_0^{t} f_0 (\tau)d \tau = f_0 t,
\end{equation} 
which is an evident simplification.
It follows that we cannot distinguish between
the frequency and phase variations in an unknown observed signal (see also \citealt{Benko2011}
for the discussion).

The uniqueness of the Fourier representation has an important consequence.
On the one hand, \citet{Benko2011} showed that the Fourier solution of a
general periodically modulated periodic signal can be written in the form of
Eq.~(\ref{mod_nsinC}). On the other hand, Fourier solution of almost periodic functions
is unique, so Eq.~(\ref{mod_nsinC2}) belongs to the
same function only if it can be transformed equivalently into Eq.~(\ref{mod_nsinC}).
But this is not possible, because Eq.~(\ref{mod_nsinC2}) represents a more
general function than Eq.~(\ref{mod_nsinC}). 

As a conclusion of this section we can state that both externally
modulated periodic signals and Blazhko light curves can also be described
by almost periodic functions but these functions are generally different.
Other words, the Blazhko effect cannot be an external modulation 
on the pulsation. Now we understand \cite{Szeidl2012} finding: they
showed, albeit they did not state directly,
that the Blazhko light curves are not modulated signals
but signals of a different and more complicated physical effect.

\section{The harmonic detuning effect}

As we have seen, the similarites of periodic and almost periodic functions have important consequences. 
This is true of their differences, as well. 
Considering the definition of the  instantaneous angular frequency 
$d\omega(t)/dt=d [2\pi f(t)] /dt$, and Eq.~(\ref{F1}),  
the instantaneous frequency of the components of an almost periodic signal is:
\begin{equation}\label{harm}
f_n(t)=n f_0 (t) + \frac{1}{2\pi} \varphi_{n}^{\prime}(t).
\end{equation}
In general, there is no one-to-one relationship between the
  instantaneous frequency and the spectral frequency in terms 
of Fourier decomposition. A detailed review of the topic can be found 
e.g. in \citet{Boashash92}. Fortunately,  the average 
frequency of the Fourier spectrum equals the time average of the instantaneous
frequency. Let us denote the time average of the quantities in Eq.~(\ref{harm}) as
$f_n=\langle f_n(t)\rangle$, $f_0=\langle f_0(t)\rangle$, and 
$\langle\varphi_n^{\prime}\rangle=\langle\varphi_n^{\prime}(t)\rangle$, and
\begin{equation}\label{harm_av}
f_n=n f_0 + \frac{1}{2\pi} \langle\varphi_{n}^{\prime}\rangle.
\end{equation}
That is, the Fourier spectrum contains not the exact harmonic
frequencies $nf_0$ but frequencies which are shifted by a term containing 
the time average of the derivative of the phase variation functions $\varphi_{n}(t)$. 
From now on this phenomenon is referred to as harmonic detuning effect (HDE).

Can this effect be detectable? The generally accepted 
picture is that the Fourier spectra of the Blazhko light curves contain the
main pulsation frequency and its exact harmonics.
In the case of V445~Lyr -- a {\it Kepler} star showing extremely complex
Blazhko effect --, however, \citet{Guggenberger2012} reported 
a `systematic and significant' deviation of the detected harmonic frequencies from 
their exact harmonic position. This finding raises the  
question whether these detected deviations are due to the HDE or not? 

\subsection{Observed frequency deviations}\label{sec:obs_dev}

\subsubsection{The case of V445\,Lyr}

To decide on this question let us investigate the reported feature in detail.
Figure 4 of \citet{Guggenberger2012} showed the detected  $f$ frequencies of V445\,Lyr
as the function of modulo $f_0$, where $f_0$ means the 
main pulsation frequency. (This $f$ modulo $f_0$ quantity is denoted here by $D$). 
If the $f_n$ components of the Fourier solution were exact harmonics ($f_n=nf_0$),
then $D(f_n)=D_n=0$ and the points which symbolise the frequencies $f_n$ would have been on a vertical
line. However, we see a slight rightward tilt of the line connecting these 
frequencies.  

The critical point is the significance of these deviations. Since there is no 
significance test in \citet{Guggenberger2012} paper, here is performed one.
If we write the deviation parameter (or shortly deviation) as 
\begin{equation}\label{Dn_def}
D_n=\left\vert \frac{f_n}{f_0} - n \right\vert,
\end{equation}
we can see that
the  $\sigma(D)$ accuracy depends only on the given
frequencies $f_0$ and $f_n$ and their accuracies
$\sigma(f_0)$ and $\sigma(f_n)$:
\begin{equation}
\sigma(D_n)=\frac{f_n\sigma(f_0)+f_0\sigma(f_n)}{f^2_0}.
\end{equation}
 So these frequencies are needed with their estimated errors
as accurately as possible.

This work used the {\it Kepler} long cadence observation of V445\,Lyr. The photometric data
were taken from the paper of \citet{Benko14}, where all characteristics of this data set were
described. It should be mentioned that, as opposed to the case of \citet{Guggenberger2012},  
now the entire  {\it Kepler} observations are available which
resulted here a $\sim$2.5 times longer (1426 days vs. 588 days) data set.  

The usual way for obtaining the frequency content of a variable star's light curve
is a consecutive pre-whitening process. This method, however, could sometimes cause 
serious troubles as well (see \citealt{Balona14}). In our particular case the problem
is that each pre-whitening step adds to the frequency uncertainty.  
To avoid this effect, the raw (unwhitened) spectrum was exclusively used.
Generally this needs extra caution, because the structure of the lower amplitude frequencies 
are dominated by the window function. Fortunately, the investigated frequencies 
are amongst the highest amplitude ones, so this problem does not affect the results.

The accuracy of the frequency determination is often estimated through the $\chi^2$ error of 
the non-linear fit of the given harmonic component (e.g. \citealt{Montgomery99, Period04}). However, 
because the present investigation does not need amplitudes and phases, the involvement of the 
amplitude and phase errors into the frequency determination could be avoided. 
A common simple frequency accuracy estimation which does not need any fits is the Rayleigh 
resolution ($\sigma_{\mathrm R}(f)=1/\Delta t$, where $\Delta t$ is the total time span of the observation). 
For the {\it Kepler} data of V445\,Lyr it is 0.0007~d$^{-1}$. 
\citet{Kallinger08} showed, however, that
the Rayleigh resolution value is a drastic overestimate of the real uncertainty and
found the value $\sigma_{\mathrm K}(f)=1/(\Delta t\sqrt{s_f}$) to be a reliable upper estimate for 
the frequency determination error. The quantity $s_f$ here the spectral significance 
of the frequency defined by \citet{Reegen07}. This error estimation 
$\sigma_{\mathrm K}(f)=\sigma(f)$ was accepted as a first approximation.
\begin{table}
        \centering
        \caption{An excerpt from the electronic table {\tt dn\_all.dat} with
the values of V445\,Lyrae. The individual columns contain the harmonic order, $n$;
the Fourier frequency, $f_n$; its error,  $\sigma(f_n)$; the  $D_n$ deviation of $f_n$
frequency from the exact harmonic; and its error,  $\sigma(D_n)$.
}
        \label{tab:mod}
        \begin{tabular}{*5{l}} 
                \hline
$n$ & $f_n$ & $\sigma(f_n)$ & $D_n$ & $\sigma(D_n)$ \\
    & (d$^{-1}$) & (d$^{-1}$) &  &  \\
                \hline
\dots &&&& \\
1 &1.9489699 &8.75E$-$06 &0.00E$+$00 &8.98E$-$06 \\
2 &3.8979614 &3.12E$-$05 &1.10E$-$05 &2.50E$-$05 \\
3 &5.8470079 &1.38E$-$04 &5.03E$-$05 &8.43E$-$05 \\
4 &7.7968746 &2.23E$-$04 &5.10E$-$04 &1.32E$-$04 \\
5 &9.7457488 &4.23E$-$04 &4.61E$-$04 &2.40E$-$04 \\
\dots &&&& \\
                \hline
        \end{tabular}
\end{table}   

The steps of the frequency finding process are as follows: 
the {\sc SigSpec} program \citep{Reegen11}
 was run once for finding the dominant frequency $f_0$ and its accuracy $\sigma (f_0)$.
Then  it was run again in a small interval around each 
$nf_0$ harmonic position ($nf_0-0.05 \le f \le nf_0+0.05$) separately 
for finding the proper position of $f_n$. Then  the highest amplitude
frequency  was double checked whether it
is really a harmonic and not a Blazhko side frequency which has higher amplitude than its
central peak. In the latter case
the best harmonic candidate was selected from the {\sc SigSpec} result file
and run the program on a narrower interval around this candidate frequency. With this process
it was achieved that all frequencies are determined in the raw (unwhitened) spectrum.
The use of the raw spectrum provides better control on the errors but at the same times it reduces
the number of significant frequencies. 
For V445\,Lyr only five harmonics have been found to be significant ($s_f$>$5$). 

The obtained frequencies, $f_n$, their calculated formal errors, $\sigma(f_n)$, 
the deviation values, $D_n$, and
their errors,  $\sigma(D_n)$ are given in the electronic table attached to this paper 
from which the values of V445\,Lyr are shown as an excerpt in Table~\ref{tab:mod}.

Plotting the deviation parameter, $D_n$, with its uncertainty, $\sigma(D_n)$, 
in the function of the harmonic order, $n$, we receive the result in the upper panel of Fig.~\ref{fig:h}.
The red filled squares show the values of V445\,Lyr (for better visibility of the trend the points
are connected). As we can see, with the increasing harmonic order the deviation also tends to increase, 
and this deviation is significant above the 4th order. With these statements the
findings of \cite{Guggenberger2012} about the systematic and significant deviation trend are validated.

\subsubsection{The {\it Kepler} sample}

\begin{figure}
        \includegraphics[width=\columnwidth]{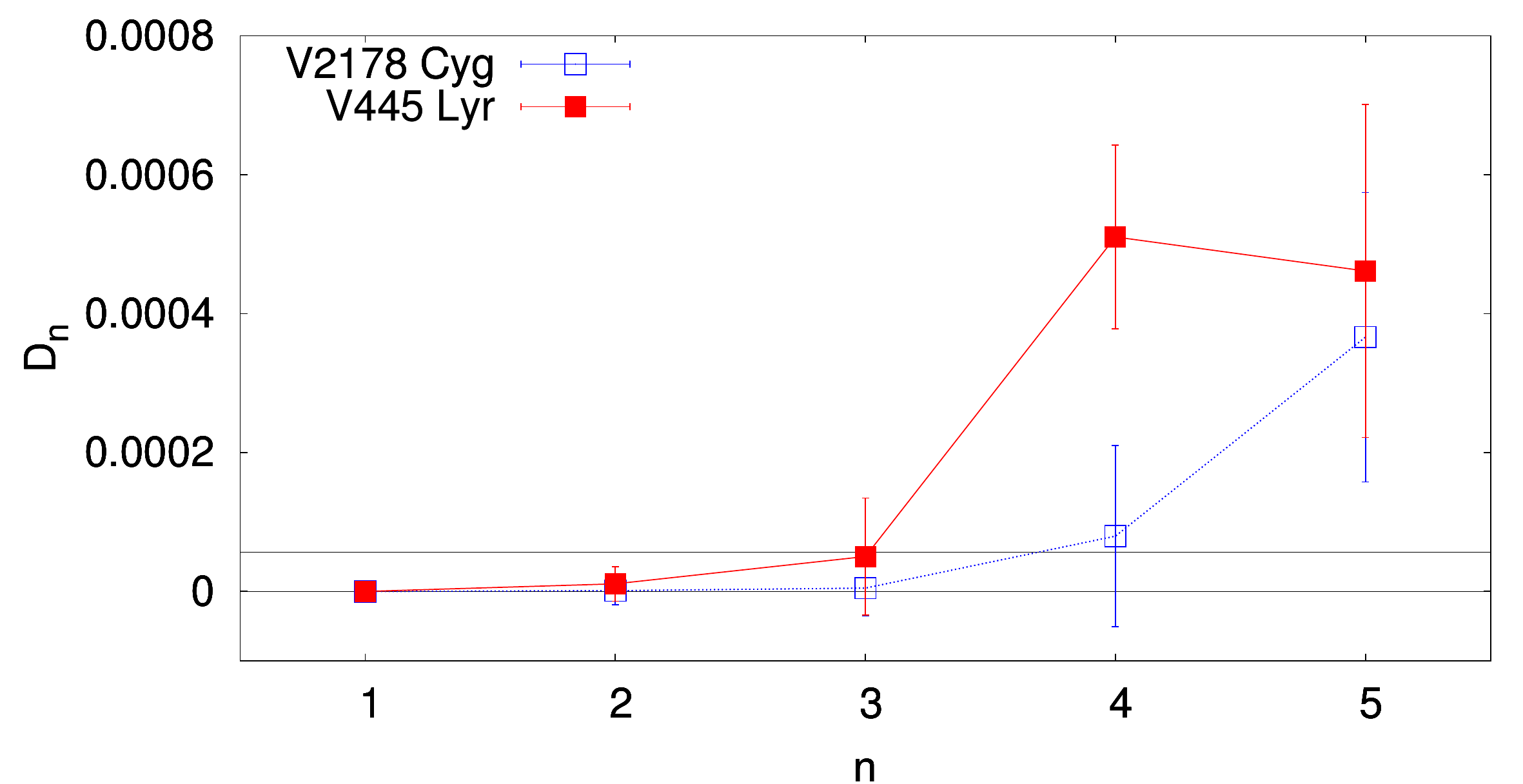}
        \includegraphics[width=\columnwidth]{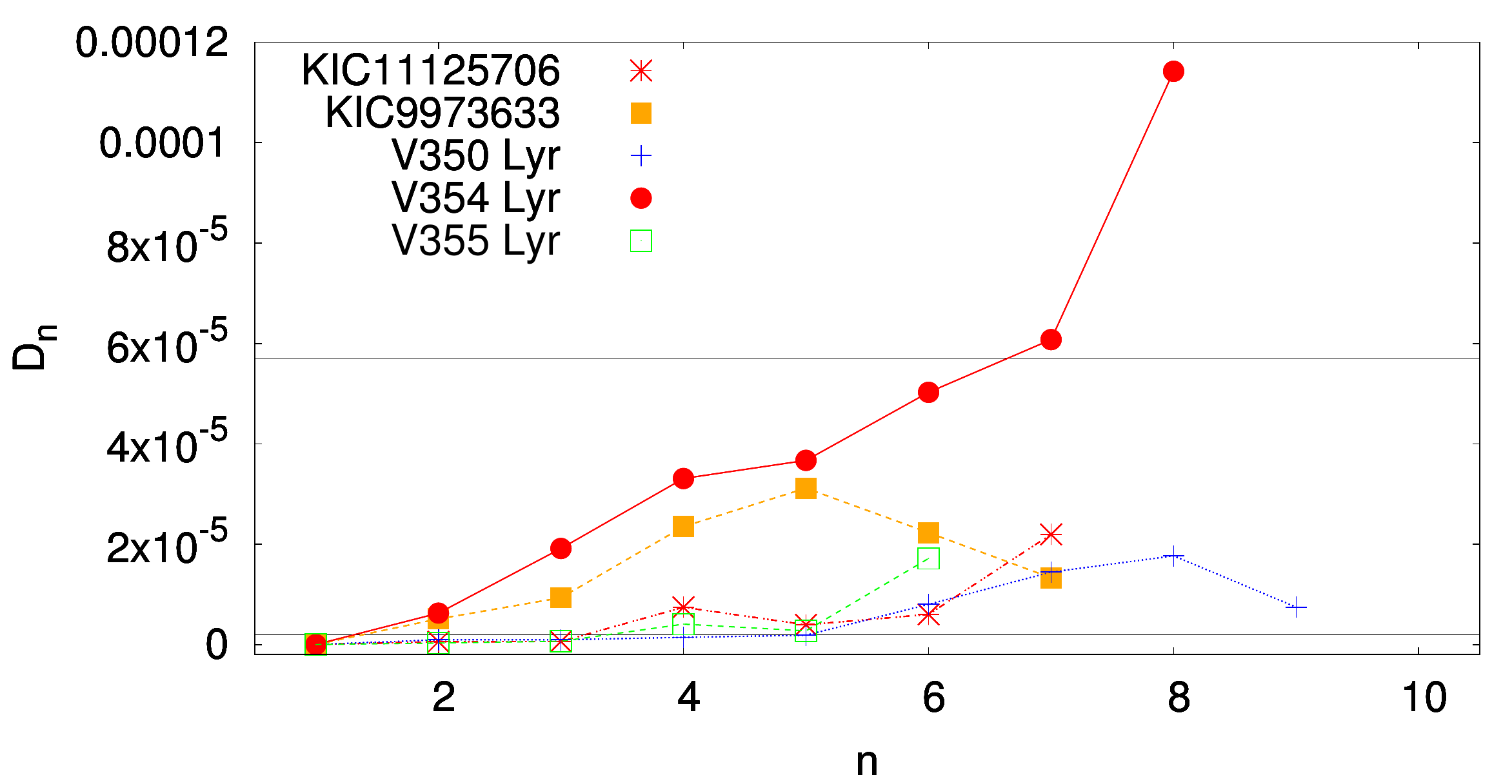}
    \caption{Deviation parameter $D_n$ as the function of the harmonic order $n$.
(top) The found two significantly detuned stars are V445\,Lyr (filled red squares) and
V2178\,Cyg (open blue squares). The black horizontal lines symbolise the range
of $D_n$ parameters of non-Blazhko stars. The formal calculated errors are shown with
the error bars.
(bottom) The next five Blazhko stars' deviation function are shown with different symbols.
The black horizontal lines are the same as the top panel. None of the plotted function
show significant deviation, but V354\,Lyr (filled red circles) shows higher deviations
than the non-Blazhko stars. For the sake of clarity the error bars are omitted here and
the points are connected with continuous lines  in both panels.
}
    \label{fig:h}
\end{figure} 
After the cited \cite{Guggenberger2012} paper no 
similar deviation trends have been reported, what is most probably, the position of Blazhko star's
harmonic frequencies have not been investigated at all. 

That is why the light curves of the fundamental mode
RR\,Lyrae (RRab) stars of the original {\it Kepler} field  were checked. The measurements of this
sample are the best ever observed in the sense of accuracy and continuous sampling. 
The quarterly stitched and tailor-made aperture photometric 
light curves were used from the papers \citet{Benko14, Benko15}.
The sample contains 17 Blazhko RRab stars (see their names in column 1 of Table~\ref{tab:sample})
which -- apart from RR\,Lyr, the prototype -- is the complete set 
of the published data. RR\,Lyr  was omitted because its overexposed
measurements would have needed special handling.

As an additional check for the accuracy of the frequency position 
the {\it Kepler} non-Blazhko RRab sample were also processed. 
Up to now 19 non-Blazhko 
stars\footnote{The present non-Blazhko sample is the same as listed by \cite{Nemec13} 
apart from those two stars (V350\,Lyr and KIC\,7021124) 
which turned out to be showing a slight Blazhko effect \citep{Benko15}.} 
were published in the original {\it Kepler} 
field \citep{Nemec11, Nemec13}. The names of the used stars are given in Table~\ref{tab:sample}.
Stitched and tailor-made aperture photometric data of non-Blazhko stars, however,
have not been available still now. For uniform handling this data set was prepared.
The {\it Kepler} long-cadence Q0-Q17 data were processed the same way as
it was described in \citet{Benko14} in details. The resulted light curves can
be downloaded from the same site as the Blazhko 
data\footnote{\url{http://www.konkoly.hu/KIK/data_en.html}}.  

The $f_n$, $\sigma(f_n)$, $D_n$, and $\sigma(D_n)$ parameters were determined for all stars. 
The applied frequency searching method was identical with the above described one
for the case of V445\,Lyrae. The result is given in the same electronic table in the same
form as it is shown for V445\,Lyr in Table~\ref{tab:mod}.  Based on the results the
following statements can be made: 

(1) Beyond the case of V445\,Lyr, significant deviation values
were found for V2178\,Cyg as well. Its $D_n$ vs. $n$ function is shown in the top panel
of Fig.~\ref{fig:h} (blue open squares). The systematic increase of the deviation 
with the increasing harmonic order is clearly visible. The deviation of the 5th component
has become significant. 

(2) No further stars of the sample show significant deviation. 
The deviation function of the five stars showing the next highest deviation parameters
are plotted in the bottom panel of Fig.~\ref{fig:h}.
Due to the short range shown in the vertical axis, indicating the errors would make the
figure obscure, the error bars are omitted. 

(3) If one selects the highest 
 $D_n$ values for each non-Blazhko star ($D_{\mathrm{max}}$) separately one found that these numbers
spread in a very limited interval between 2$\cdot10^{-6}$ and 5.7$\cdot10^{-5}$
(see column 4 in Table~\ref{tab:sample}).
These two limit values are represented by horizontal lines in the panels of  Fig.~\ref{fig:h}.
As we see the Blazho stars' values are within this range except for V354\,Lyr. 

(4) If we compare the measured $D_{\mathrm{max}}$ values in column 4 of Table~\ref{tab:sample}
with the formal errors $\sigma(D_{\mathrm{max}})$ (column 3)
calculated from the $\sigma_{\mathrm K}$ values (in column 2), one finds that 
this calculated formal error might be a 5-10 times overestimation of the measured deviations.
In the light of this, V354\,Lyr is a possible candidate star for
showing HDE.
 
\begin{table}
        \centering
        \caption{The used {\it Kepler} RR Lyrae sample.
The columns show the star's name; the maximal value of the estimated frequency error
 of the star's harmonic components; the highest calculated deviation 
error; the maximum value of the deviation.}
        \label{tab:sample}
        \begin{tabular}{r*5{c}} 
                \hline
Name/KIC & $\sigma_{\mathrm {K,max}}$ 
&  $\sigma(D)_{\mathrm{max}}$ & $D_{\mathrm{max}}$ \\
& ($\times 10^{-4}$d$^{-1}$) 
&  ($\times 10^{-4}$) & ($\times 10^{-4}$) \\
                \hline
\multicolumn{4}{c}{Blazhko stars}  \\
\hline
V783\,Cyg & 1.89 & 1.49 & 0.04 \\
V808\,Cyg & 4.90 & 3.01 & 0.12 \\
V838\,Cyg & 2.83 & 1.71 & 0.16 \\
V1104\,Cyg & 2.59 & 1.42 & 0.03 \\
V2178\,Cyg & 3.88 & 2.09 & 3.66 \\
V350\,Lyr & 2.42 & 1.82 & 0.18  \\
V353\,Lyr & 2.15 & 1.48 & 0.12 \\
V354\,Lyr & 4.02 & 2.63 & 1.14 \\
V355\,Lyr & 1.90 & 1.11 & 0.17 \\
V360\,Lyr & 2.40 & 1.57 & 0.03 \\
V366\,Lyr & 2.27 & 1.46 & 0.08  \\
V445\,Lyr & 4.23 & 2.40 & 5.10 \\
V450\,Lyr & 1.88 & 1.16 & 0.07 \\
7021124 & 2.56 & 2.00 & 0.03 \\
7257008 & 4.99 & 3.29 & 0.07 \\
9973633 & 6.14 & 4.16 & 0.31 \\
11125706& 2.93 & 2.15 & 0.22 \\
\hline
\multicolumn{4}{c}{non-Blazhko stars}  \\
\hline
AW\,Dra & 2.92 & 2.44 & 0.31 \\
V715\,Cyg & 2.70 & 1.61 & 0.13 \\
V782\,Cyg & 2.95 & 1.80 & 0.09 \\
V784\,Cyg & 2.49 & 1.55 & 0.02  \\
V839\,Cyg & 4.15 & 2.38 & 0.34 \\
V894\,Cyg & 2.87 & 2.08 & 0.22 \\
V1107\,Cyg & 2.20 & 1.52 & 0.04 \\
V1510\,Cyg & 2.95 & 2.13 & 0.15 \\
V2470\,Cyg & 2.44 & 1.60 & 0.33 \\
FN\,Lyr & 2.14 & 1.57 & 0.46  \\
NQ\,Lyr & 2.61 & 1.87 & 0.26 \\
NR\,Lyr & 2.42 & 2.09 & 0.57 \\
V346\,Lyr & 2.74 & 1.95 & 0.22 \\
V349\,Lyr & 1.98 & 1.29 & 0.12 \\
V368\,Lyr & 2.63 & 1.49 & 0.06 \\
6100702 & 3.04 & 1.72 & 0.06 \\
7030715 & 2.87 & 2.34 & 0.39 \\
9658012 & 5.52 & 3.78 & 0.33 \\
9717032 & 6.02 & 4.22 & 0.28 \\
                \hline
        \end{tabular}
\end{table}

\subsubsection{Alternate explanation for the deviations}

In the previous subsections significant harmonic
detunig was detected for at least two {\it Kepler} Blazhko RR Lyrae stars.
Since the light curve description using almost periodic functions 
directly predicts such an effect (the HDE) 
for the detected deviations HDE seems to be the most plausible explanation. 

The paper of \cite{Guggenberger2012} gives two possible explanations for the
detected deviations of V445\,Lyr: (1) `result of period changes', or
(2) `close unresolved peaks' in the Fourier spectrum. The first case is 
closely related to the HDE explanation since the period change means
phase variation and the phase variation functions play the key role in HDE
(see Eq.~\ref{harm_av}). 
The second option implicitly assumes the existence of a very long period (maybe secular)
or a nearly pulsation period change that would imply an unresolved close frequency 
$f_{\mathrm c}$. The other `unresolved' frequencies would
be the linear combinations of $f_{\mathrm c}$ with the harmonics of the 
main pulsation frequency. Because of their identical frequency differences 
the observed different $D_n$ values would be difficult to interpret.

The case of V354 Lyr raises a third possibility.
The $D_n$ values of V354 Lyr fit well to a linear trend up to $n=7$ 
with the slope of $1.03\cdot10^{-5}$. This can be explained as follows.
The frequencies
are always inaccurately known: $f^{\delta}_0=f_0+\delta_0$, $f^{\delta}_n=f_n+\delta_n$.
(Here the measured inaccurate frequencies are upper indexed by a $\delta$, 
their errors are denoted by $\delta_0$ and $\delta_n$,
the exact quantities are $f_0$ and $f_n$.)
If $\delta_n = n(\delta_0 + \delta^{\prime})$ ($\delta^{\prime}\ne 0$),
the $D_n$ value increases linearly with the harmonic order, $n$, even 
if the harmonic frequencies are in the exact harmonic position ($f_n=nf_0$).
This argument assumes that the error of the harmonic frequencies is 
sytematically different from the error of the main frequency 
($\delta_n\ne n\delta_0$). This cannot be ruled out completely even though we have 
done everything to avoid it.

If we could get compatible numerical values from the observed material
and the prediction of HDE, 
it would clearly demonstrate that the measured differences were caused by HDE. 
We will try this in the next section.

\subsection{Phase variation function}

\begin{figure*}
        \includegraphics[width=\textwidth]{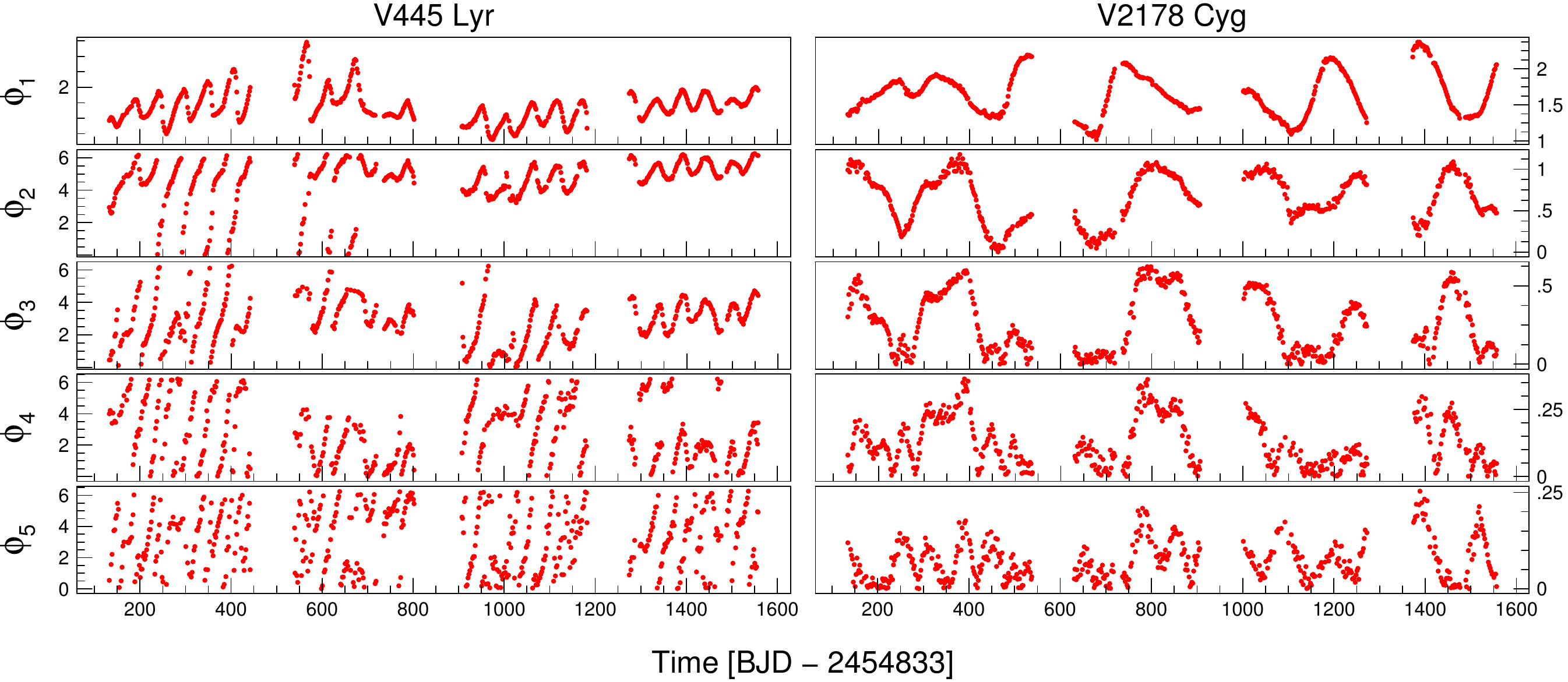}
    \caption{The first five  Fourier phase variation function $\varphi_n(t)$ ($n=1,2, \dots 5$) 
for the {\it Kepler} Blazhko stars V445\,Lyr and V2178\,Cyg. 
}
    \label{fig:phase_var}
\end{figure*} 
In Sec.~\ref{sec:obs_dev} the 
deviation parameter, $D_n$, was measured with the formula (\ref{Dn_def}). 
We can express this quantity by an alternate form as well.
From Eq.~(\ref{harm_av}) and (\ref{Dn_def})
\begin{equation}\label{eq:dn} 
D_n=\left\vert \frac{\langle\varphi_n^{\prime}\rangle}{2\pi f_0} \right\vert. 
\end{equation}
If we determine the $D_n$ values from both  (\ref{Dn_def}) and (\ref{eq:dn})
and find that these are compatible values then 
we would justify formula~(\ref{harm_av}) which would make 
the almost periodic explanation more robust. 

Expression~(\ref{eq:dn}) includes the main pulsation frequency $f_0$ as it can be determined
from the spectrum and the time average of the derivative of the phase
variation function. The latter parameter has never been determined for any stars.  
The temporal variability of the Fourier phases of the Blazhko stars' harmonic
frequencies were discovered and studied already from the best
ground-based observations (e.g. \citealt{Jurcsik05, Jurcsik06, Jurcsik08})
but in these cases the phase variations were typified by Fourier phases vs. Blazhko
phase functions.  The first phase-time functions could only be prepared in the space 
photometry era \citep{Guggenberger11, Guggenberger2012, Nemec11, Nemec13}. 

There are two obvious ways of determining $\langle\varphi_n^{\prime}\rangle$ values.
(1) Using the definition of the time average of a function:
\begin{equation}\label{eq:av}
\langle\varphi_n^{\prime}\rangle = \frac{1}{\Delta t}\int_{t_0}^{t_{\mathrm{l}}} \varphi_n^{\prime}(t)dt = 
\frac{1}{\Delta t} \left[ \varphi_n(t_{\mathrm{l}})-\varphi_n(t_0) \right],
\end{equation}
where $t_0$, $t_{\mathrm{l}}$, and $\Delta t$ mean the starting and ending epochs, and
the total time span of the observation, respectively. (2) The alternate possibility is to 
determine the $\varphi^{\prime}_n(t)$ functions 
(by a numerical derivation of the $\varphi_n(t)$ functions) then to average them.
It is well known that the phases are the least accurately determinable
Fourier parameters in any Fourier fits. 
Unfortunately both methods need the $\varphi_n(t)$ functions. 
Moreover, the second method requires the numerical
derivation as well, which causes additional numerical inaccuracies, 
so the first one  has been chosen. 

The $\varphi_n(t)$ functions  were determined for the two stars 
which show the HDE (V445\,Lyr and V2178\,Cyg).
The  $\varphi_n(t)$ functions were prepared with the phase variation calculation
tool of the {\sc Period04} \citep{Period04} package. This tool subdivides
the total observed time span into short time intervals (bins) then it applies
a non-linear fit for each bin separately. 
For compatibility, in the case of V445\,Lyr, 
the same Fourier solution ($f_0$ and its ten harmonics) and  bin size (2 days) was used
as \citet{Guggenberger2012}. As they mentioned, this bin size is optimal 
because in such a short time-scale the impact of the Blazhko effect is negligible.
So this work used the 2-d bin size for V2178\,Cyg as well. In the latter case
the used Fourier fit contains $f_0$ and its 12 harmonics. 
The number of harmonics here is higher than  we found to be significant 
in Sec.~\ref{sec:obs_dev}, but now the goal is to calculate the $\varphi_n(t)$ functions 
as  accurately as possible, and therefore all harmonics from the conventional pre-whitening
process should be considered.

The  first five $\varphi_n(t)$ functions ($n=1,2,\dots,5$) of V445\,Lyr and V2178\,Cyg 
are shown in Fig.~ \ref{fig:phase_var}.
The functions of V445\,Lyr for $n>1$ show long time segments, 
where the phase continuously increases exceeding the $(0, 2\pi)$ interval. 
The summed length of such time intervals increases with increasing $n$.
The functions of V2178\,Cyg do not show this feature, what is more, their
amplitudes tend to decrease with the increasing harmonic order. 
It must be mentioned that the
scatter shown in the figures is, partly, because of an intrinsic effect: the signal
of the excited low amplitude modes present in the stars (see \citealt{Guggenberger2012, Benko14})
disturb the light curves on short time-scale causing higher scatter.  

By applying Eq.~(\ref{eq:dn}) and (\ref{eq:av}) we obtain for the 
deviation parameters
$D_4=0.33\pm9.3\cdot10^{-4}$,
$D_5=0.93\pm3.6\cdot10^{-4}$ for  V445\,Lyr, 
and 
$D_5=0.62\pm7.5\cdot10^{-5}$ for V2178\,Cyg.
First, we see that estimated errors are quite large, as we predicted.
Second, these values are systematically lower than the values 
obtained in Sec.~\ref{sec:obs_dev} (see in Table~\ref{tab:mod} and Table~ \ref{tab:sample}).
If we consider the errors as well, the values become consistent within 3$\sigma$ intervals, but 
because of the relatively large errors, all we can state is that the calculated deviation values
of this section do not contradict the results of Sec.~\ref{sec:obs_dev}.

Although formula~(\ref{eq:dn}) proved to be not suitable for calculating deviation parameter,
it gives ideas when the HDE becomes detectable: shortly, if $\langle\varphi^{\prime}_n\rangle$ has
a relatively large, non-zero value. The $\varphi(t)$ functions of the Blazhko stars 
(and consequently their derivatives $\varphi^{\prime}_n(t)$ as well)
are generally nearly periodic with the Blazhko period(s). Let us assume the phase 
variation functions to be strictly periodic and represent them with their Fourier sum:
\begin{equation}\label{eq:phi_n}
\varphi_n(t)=\frac{a^{\varphi}_{n0}}{2} +
\sum_{i=1}^{\infty}{a^{\varphi}_{ni} \sin \left( 2\pi i f_{\mathrm B} t + \varphi^{\varphi}_{ni} \right)},
\end{equation}
where $a^{\varphi}_{n0}$, $a^{\varphi}_{ni}$, $\varphi^{\varphi}_{ni}$ are the 
constant Fourier coefficients, and $f_{\mathrm B}$ is the Blazhko frequency. 
Using (\ref{eq:av}) and some trigonometric identities we obtain the trivial result:
\begin{multline}\label{eq:phi_n_av}
\langle\varphi_n^{\prime}\rangle = \frac{2}{\Delta t}\sum_{i=0}^{\infty} 
a^{\varphi}_{ni}\cos \left[ \pi i f_{\mathrm B}(2t_0+\Delta t) + \varphi^{\varphi}_{ni} \right]\times \\
\sin \left( \pi i f_{\mathrm B}\Delta t \right) = 0.
\end{multline}
That is, in the case of fully periodic phase variation functions the deviation 
parameter is zero. As the phase variation functions differ from periodic ones,
$D_n$ became non-zero. If we find a Blazhko star with highly non-periodic
Blazhko effect (which causes non-periodic phase variation functions) we expect 
the HDE to appear.

The O$-$C diagram well characterises the frequency variation part of the 
Blazhko effect. It contains the total frequency (or phase) variation (not the
individual components), but if we see the O$-$C diagrams of the {\it Kepler} 
sample in Fig.~7 of \citet{Benko14} we find that V445\,Lyr and V2178\,Cyg
show the two largest amplitude and most exotic (non-periodic) diagrams.
In other words, checking the phase variation or O$-$C curves could help us 
to select those stars which might show detectable HDE.

\section{Conclusions}

In this short paper it is shown that two common assumptions concerning RR\,Lyrae
light curves seems to be invalid.

$\bullet$
(1) 
It was shown that the mathematical formalism currently best 
describing the observed Blazhko light curves  \citep{Szeidl2012}
is the Fourier representation of general almost periodic functions.  
It follows that, the widely used phrase `Blazhko modulation' is not precise.
Although many features of the Blazhko light curves can be described in the modulation 
framework, the real light curves are more than a simple modulated signals.  
That is, the external modulation explanation of the Blazhko effect is deficient, which
gives a further support for those theoretical models which produce the effect 
 as an inherent feature of the pulsation (e.g. \citealt{Buchler2011, Kollath16}).

$\bullet$
(2) Up to now most studies assumed that the Fourier spectra of the light curve of
an RR Lyrae star, either mono-periodic or showing the Blazhko effect, 
is dominated by the main pulsation frequency and its harmonics. 
The Fourier representation of almost periodic functions, however, does
not consist of the harmonics of the main pulsation frequency, but
terms with frequencies which are slightly different from the exact harmonics.
This was called as harmonic detuning effect (HDE).

By investigating the {\it Kepler} Blazhko RR Lyrae sample concerning the HDE
the following conclusions  can be drawn:

$\bullet$
On the one hand, the previously found deviation of the 
harmonic frequencies in the Fourier spectrum of V445\,Lyr can be explained
with HDE. On the other hand,
an additional star -- V2178\,Cyg --  was found showing significant HDE. 
The detected deviation of V354\,Lyr is formally non-significant, 
 but unusually high compared to other Blazhko and non-Blazhko stars.  
That is why V354\,Lyr can be considered a candidate HDE showing star.

$\bullet$
As a by-product of this work, 
stitched tailor-made aperture light curves for the 19 {\it Kepler} non-Blazhko RRab stars
were prepared and made publicly available\footnote{\url{http://www.konkoly.hu/KIK/data_en.html}}.

$\bullet$
The HDE can be detected for those  
Blazhko stars only which have (i) high amplitude phase variation
-- otherwise the deviations are below the detection limit.
(ii) These phase variations should also be non-periodic enough
-- otherwise the effect is averaged out. There is a third condition for
the detectability of HDE: (iii) the Blazhko light curves should
be long and precise enough -- otherwise the error of frequency 
determination makes this small effect undetectable.
These strong conditions explain why HDE was not previously revealed.

\section*{Acknowledgements}

The author thanks to R. Szab\'o for reading and to L. Sza\-ba\-dos for language editing
the manuscript, likewise thanks to the anonymous referee for
her/his valuable comments. This project has been supported by the Hungarian National Research, 
Development and Innovation Office -- NKFIH K-115709.





\begin{thebibliography}{99}
\bibitem[Baglin et al.(2006)]{Baglin06}
Baglin, A. Auvergne, M., Boisnard, L. et al. 2006, in 
36th COSPAR Scientific Assembly, ed. A. Wilson, ESA SP 1296, (ESA, Noordwijk), 3749
\bibitem[\protect\citeauthoryear{Balona}{2014}]{Balona14}
Balona, L.~A. 2014, \mnras, 439, 3453
\bibitem[\protect\citeauthoryear{Benk\H{o} \& Papar\'o}{2013}]{Hakone_conf}
Benk\H{o}, J.~M., Papar\'o, M. 2013, in Progress in Physics of the Sun and Stars: 
A New Era in Helio- and Asteroseismology, eds. Shibahashi, H. \& Lynas-Gray, A.~E.,
ASP Conf. Ser. 479, 531
\bibitem[\protect\citeauthoryear{Benk\H{o} \& Szab\'o}{2015}]{Benko15}
Benk\H{o}, J.~M., Szab\'o, R. 2015, \apjl, 809, L19
\bibitem[\protect\citeauthoryear{Benk\H{o} \& Szab\'o}{2016}]{Visegrad_conf}
Benk\H{o}, J.~M., Szab\'o, R. 2016,
 in RRL2015: High-Precision Studies of RR~Lyrae Stars, 
eds. Szabados, L., Szab\'o, R. \& Kinemuchi, K., Comm. Konkoly. Obs. No.~105, p. 197
\bibitem[\protect\citeauthoryear{Benk\H{o} et al.}{2010}]{Benko10}
Benk\H{o}, J.~M. et al. 2010, \mnras,  409, 1585 
\bibitem[\protect\citeauthoryear{Benk\H{o} et al.}{2011}]{Benko2011}
Benk\H{o}, J.~M., Szab\'o, R., Papar\'o, M. 2011, \mnras, 417, 974 
\bibitem[\protect\citeauthoryear{Benk\H{o} et al.}{2014}]{Benko14}
Benk\H{o}, J.~M., Plachy, E., Szab\'o, R., Moln\'ar, L., Koll\'ath, Z. 2014, \apjs, 213, id.31
\bibitem[\protect\citeauthoryear{Benk\H{o} et al.}{2016}]{Benko16}
Benk\H{o}, J.~M., Szab\'o, R., Derekas, A.,  S\'odor, \'A. 2016, \mnras,  463, 1769
\bibitem[\protect\citeauthoryear{Boashash}{1992}]{Boashash92}
Boashash, B. 1992, Proc. of the IEEE, 80, 520
\bibitem[\protect\citeauthoryear{Bohr}{1947}]{Bohr1947}
Bohr, H. 1947, Almost Periodic Functions, Chelsea, New York 
\bibitem[Borucki et al.(2010)]{Borucki10}
Borucki, W.~J. et al. 2010, \sci, 327, 977
\bibitem[\protect\citeauthoryear{Brown et al.}{2011}]{KIC}
Brown, T.~M., Latham, D.~W., Everett, M.~E.,  Esquerdo, G.~A. 2011, \aj, 142, 112 
\bibitem[\protect\citeauthoryear{Buchler \& Koll\'ath}{2011}]{Buchler2011}
Buchler, J.~R., Koll\'ath, Z. 2011, \apj, 731, id.24 
\bibitem[\protect\citeauthoryear{Guggenberger et al.}{2011}]{Guggenberger11}
Guggenberger E., Kolenberg, K., Chapellier, E.,  Poretti, E., Sza\-b\'o, R., 
Benk\H{o}, J.~M., Papar\'o, M. 2011, \mnras, 415, 1577 
\bibitem[\protect\citeauthoryear{Guggenberger et al.}{2012}]{Guggenberger2012}
Guggenberger E. et al. 2012, \mnras, 424, 649
\bibitem[\protect\citeauthoryear{Jurcsik et al.}{2005}]{Jurcsik05}
Jurcsik J. et al. 2005, \aap, 430, 1049 
\bibitem[\protect\citeauthoryear{Jurcsik et al.}{2006}]{Jurcsik06}
Jurcsik J. et al. 2006, \aj, 132, 61 
\bibitem[\protect\citeauthoryear{Jurcsik et al.}{2008}]{Jurcsik08}
Jurcsik J. et al. 2008, \mnras, 391, 164 
\bibitem[\protect\citeauthoryear{Kallinger, Reegen \& Weiss}{2008}]{Kallinger08}
Kallinger, T., Reegen, P., Weiss, W.~W. 2008, \aap, 481, 571 
\bibitem[\protect\citeauthoryear{Kolenberg et al.}{2006}]{Kolenberg06} 
Kolenberg, K. et al. 2006, \aap, 459, 577  
\bibitem[\protect\citeauthoryear{Koll\'ath}{2016}]{Kollath16}
Koll\'ath, Z. 2016, in RRL2015: High-Precision Studies of RR~Lyrae Stars,
eds. Szabados, L., Szab\'o, R. \& Kinemuchi, K.,
Comm. Konkoly Obs. No.~105, p. 35
\bibitem[\protect\citeauthoryear{Lenz \& Breger}{2005}]{Period04}
Lenz, P., Breger, M. 2005, CoAst, 146, 53
\bibitem[\protect\citeauthoryear{Montgomery \& O'Donoghue}{1999}]{Montgomery99}
Montgomery, M.~H., O'Donoghue, D., 1999, Delta Scuti Star Newsletter, 13, 28 
\bibitem[\protect\citeauthoryear{Nemec et al.}{2011}]{Nemec11}
Nemec, J.~M. et al. 2011, \mnras, 417, 1022
\bibitem[\protect\citeauthoryear{Nemec et al.}{2013}]{Nemec13}
Nemec, J.~M., Cohen, J.~G., Ripepi, V., Derekas, A., Moskalik, P., Sesar, B., Chadid, M., 
Bruntt, H. 2013, \apj, 773, id.181
\bibitem[\protect\citeauthoryear{Reegen}{2007}]{Reegen07}
Reegen, P. 2007, \aap, 467, 1353
\bibitem[\protect\citeauthoryear{Reegen}{2011}]{Reegen11}
Reegen, P. 2011, CoAst, 163, 3
\bibitem[\protect\citeauthoryear{Szab\'o et al.}{2014}]{Szabo14}
Szab\'o, R. et al.  2014, \aap, 570, A100  
\bibitem[\protect\citeauthoryear{Szeidl et al.}{2012}]{Szeidl2012}
Szeidl B., Jurcsik, J., S\'odor, \'A., Hajdu, G.,  Smitola, P. 2012, \mnras, 424, 3094
\end{thebibliography}



\bsp	
\label{lastpage}

\end{document}